\documentclass[12pt]{iopart}
\usepackage{graphicx}
\usepackage{iopams}

\newcommand*{\Ai}{\mathop{\textrm{Ai}}\nolimits}
\newcommand*{\Bi}{\mathop{\textrm{Bi}}\nolimits}

\newcommand*{\sgn}{\mathop{\textrm{sgn}}\nolimits}

\def\pd#1#2{\frac{\partial #2}{\partial #1}}   

\newcommand{\abs}[1]{\left\vert#1\right\vert}

\begin{document}
\title[WKB for a ceiling]{WKB propagators in position and momentum
space for a linear potential with a ``ceiling'' boundary}
\author{T A Zapata$^{1,3}$ and S A Fulling$^{1,2}$}
\address{$^1$ Department of Physics, Texas A\&M University, 
College Station, Texas, USA}
\address{$^2$ Department of Mathematics, Texas A\&M University, 
College Station, Texas, 77843-3368, USA}
\address{$^3$ Present address: Department of Electrical and 
Computer Engineering, Texas A\&M University, College Station, 
Texas, 77843-3128, USA}
\eads{\mailto{todd-austin-zapata@tamu.edu}, 
\mailto{fulling@math.tamu.edu}}
\begin{abstract}
As a model for the semiclassical analysis of quantum-mechanical 
systems with both potentials and boundary conditions, we 
construct the WKB propagator for a linear potential sloping away 
from an impenetrable boundary.  First, we find all classical 
paths from point $y$ to point $x$ in time $t$ and calculate the 
corresponding action and amplitude functions. A large part of 
space-time turns out to be classically inaccessible, and the 
boundary of this region is a caustic of an unusual type, where 
the amplitude vanishes instead of diverging. We show that this 
curve is the limit of caustics in the usual sense when the 
reflecting boundary is approximated by steeply rising smooth 
potentials. Then, to improve the WKB approximation we construct 
the propagator for initial data in momentum space; this requires 
classifying the interesting variety of classical paths with 
initial momentum $p$ arriving at $x$ after time~$t$. The two 
approximate propagators are compared by applying them to Gaussian 
initial packets by numerical integration; the results show 
physically expected behavior, with advantages to the 
momentum-based propagator in the classically forbidden regime 
(large~$t$).
\end{abstract}
\maketitle

\section{\label{Intro}Introduction}

The propagator of a quantum-mechanical system is the integral kernel 
(Green function)
that solves the initial-value problem for its time-dependent 
Schr\"odinger equation:
\begin{equation}
\rmi \hbar\,\pd{t}{\psi} = H \psi
\mathrel{\Longleftrightarrow} \psi(\mathbf{x},t) = 
\int U(\mathbf{x},\mathbf{y},t)
\psi(\mathbf{y},0)\, \rmd\mathbf{y}.
\label{genprop}\end{equation} 
The semiclassical or WKB ansatz, in its simplest form, is to write 
\begin{equation}
U(\mathbf{x},\mathbf{y},t)
 = A(\mathbf{d}_\mathbf{y})\rme^{\rmi S(\mathbf{d}_\mathbf{y})/\hbar}
\qquad (\mathbf{d}_\mathbf{y} \equiv [(\mathbf{y},0),(\mathbf{x},t)])
\label{genWKB}\end{equation}
and to choose $S((\mathbf{y},s),(\mathbf{x},t))$ and 
$A((\mathbf{y},s),(\mathbf{x},t))$
so that \eref{genprop} is satisfied up to order $\hbar^2$.  
The \emph{action} $S$ and \emph{amplitude} $A$
 can be constructed by solving the classical equations of motion 
and integrating 
certain first-order differential equations along the resulting 
paths (trajectories) in space-time.

The classical-mechanical problem that must be solved in this 
construction is a two-point boundary-value problem, since the 
initial point $\mathbf{y}$, the final point $\mathbf{x}$, and the 
elapsed time $t-s$ (for time-independent Hamiltonians) are 
prescribed.  This problem is generally not well-posed, in the 
sense that for certain data there may be more than one path, for 
others no path at all (``classically forbidden regions''), and at 
still other points $A$ may become infinite. This last phenomenon 
is called a \emph{caustic}, which technically can be defined as a 
set where the mapping to $\mathbf{x}$ from the initial momentum 
$\mathbf{p}$ (at~$\mathbf{y}$) fails to be a diffeomorphism. 
Ordinarily a caustic represents a breakdown of the WKB 
approximation (although the WKB propagator of a harmonic 
oscillator is exact, despite having caustics). A semiclassical 
solution that is accurate in the vicinity of a caustic often can 
be found by transforming, at least temporarily, into a momentum 
representation \cite{MF,delos,litjohn}.

The literature of semiclassical approximation is well developed 
for Hamiltonians $H = \mathbf{p}^2/2m + V(\mathbf{x})$ (and much 
more general phase-space functions) in $\mathbf{R}^n$ without 
boundaries, as considered in the classic paper of Van\,Vleck 
\cite{vv} and many other works, such as \cite{MF,delos,litjohn}. 
The case $H= -\nabla^2$, with a reflecting boundary but no 
potential, is also well studied (for example in \cite{KR}, or 
indeed in all geometrical optics). But there has been little or 
no attention to problems with both boundaries and potentials, 
although the small-$t$ asymptotic expansion of $U$, or rather its 
close relative the heat kernel, is well known for very general 
operators with boundaries, potentials, Riemannian curvature and 
external gauge fields \cite{BGil,kirsten}.

In this paper we construct in detail two WKB propagators for a 
simple system: The configuration space is one-dimensional, the 
positive $x$~axis; the potential is a decreasing, linear function 
of~$x$; the boundary condition at the origin is perfect 
reflection (Dirichlet). Because classically this potential 
describes a force pulling a particle away from the boundary, we 
think of the force as gravitational and call this boundary a 
\emph{ceiling}. (Of course, it could at least equally well be a 
constant electrostatic force.
 The problem also arises in the study of diffraction by a smooth convex
 obstacle \cite{SS}.) The extension to dimension 3 with a flat 
ceiling is immediate, since the propagator factors and the WKB 
approximation for the transverse propagator is exact. This model 
already presents many of the features of the general case: Since 
space-time is two-dimensional, the full multidimensional WKB 
theory (using tools from Hamilton--Jacobi theory) is needed. The 
WKB approximation is not exact for the wave reflected from the 
ceiling, so this model is not ``trivial'' in the sense that the 
WKB propagator for a linear or quadratic potential in all of 
$\mathbf{R}^n$~is. The spectrum is continuous, which complicates 
the solution of the Schr\"odinger equation by separation of 
variables; on the other hand, no classical path strikes the 
ceiling more than once, so the WKB solution is relatively simple. 
(The spectrum is also unbounded below, but only because of what 
happens as $x\to+\infty$, so no physical pathology results.) In 
the semiclassical construction of $U(x,y,t)$ a certain region of 
space-time is classically forbidden, because if $t$ is too large 
compared to $x$ and~$y$, ``a ball cannot be thrown from $y$ to 
$x$ in exact time $t$ because the ceiling gets in the 
way''\negthinspace. The boundary of that region is not a caustic 
in the usual sense, because the amplitudes do not diverge there; 
however, the amplitude of the wave reflected from the ceiling 
does go to $0$ there, indicating a nondiffeomorphic dynamics, and 
its derivatives do become singular there. To investigate 
semiclassically what happens on and beyond this critical curve, 
we construct the propagator $U(x,p,t)$ that yields, in analogy to 
\eref{genprop}, the wave function $\psi(x,t)$ in terms of the 
initial wave function in momentum space, $\hat\psi(p,0)$. (This 
step is the paper's most original contribution to the 
literature.) The momentum-based calculation
 seems to be superior to the position-based one,
at least for certain initial data and times, 
 because it does not falsely predict that the propagator is 
identically $0$ at large~$t$.

In the remainder of this introduction we set up notation and 
review some basic formalism. In \sref{position} we find all 
solutions of the classical two-point boundary value problem for 
initial position~$y$ (and given destination $(x,t)$).  If $t$ is 
sufficiently small, there are always two solutions, one that 
never reaches the ceiling and one that bounces off it.  If $t$ is 
too large, there are no solutions.  There is a unique solution on 
the \emph{critical curve}, \eref{criticalbounce}. We calculate 
the action and amplitude for both the direct and the bounce path, 
thereby obtaining an approximate propagator $U(x,y,t)$ adequate 
for points inside (and not too near) the critical curve. In 
\sref{caustic} we elucidate the critical curve by replacing the 
ceiling by a smooth but rapidly rising potential.  Numerical 
solutions for families of classical paths reveal caustics (in the 
familiar sense) that converge, in the limit of an infinitely 
steep potential, to the union of the critical curve and the 
portion of the ceiling that gets struck by the bounce paths. In 
\sref{momentum} we solve the two-point boundary value problem for 
initial momentum $p$ in place of~$y$; in this case the structure 
of the set of allowed paths is more complicated.  Again there is 
a mild caustic-like behavior (compare \eref{Apb2} with 
\eref{Aby}), but it has moved to a different part of phase space. 
Again the actions and amplitudes are calculated. The resulting 
approximate propagator $U(x,p,t)$, when applied to an initial 
wave packet concentrated at values of $y$ for which $(x,t)$ would 
be in the classically forbidden region or on the critical curve, 
gives a plausible, nonvanishing value to the wave function. We 
have not proved error bounds, but in \sref{numerical} we offer 
some preliminary numerical calculations on Gaussian initial 
packets that show the expected physical behaviors.

Although an exact analytical solution for the propagator 
$U(x,y,t)$ exists as an integral of Airy functions (see 
\sref{eigen}), it is difficult (for the present authors, at 
least) to use in numerical work, so we have no trustworthy 
comparisons to present.  In contrast, the related problem with a 
\emph{floor} instead of a ceiling has been often studied (e.g., 
\cite{gea,geacom1,geacom2}); it has a discrete spectrum. In that 
case it is the WKB approach that runs into complications, because 
there are infinitely many classical paths to sum over.

  The detailed solution of the ceiling problem with linear 
potential, presented in this paper, points the way to a 
prescription for handling boundaries in problems with arbitrary 
potentials (and perhaps ultimately also higher-dimensional 
problems with curved boundaries).
 This ``localization'' strategy is analogous to the treatment of 
diffraction from corners by applying the formulas for diffraction 
by an infinite straight wedge \cite{Keller}, and it provided the 
original motivation for this work.

This  paper, especially \sref{momentum} and \sref{numerical}, 
is based on the Master of Science thesis of the first author 
\cite{zapata}.
 A  preliminary account of the conclusions in \sref{position} and
 \sref{caustic} was presented by the second author at QMath8,
 Taxco, Mexico, December 2001. 

\subsection{The classical system}  \label{classicsys}
The system considered in this article is a 1-dimensional one
with an impenetrable
barrier at the origin and a linear potential,
\begin{eqnarray}
V(q)=-\alpha q.  
\label{potential}\end{eqnarray}
Here $q$ represents the position of the particle, and $\alpha$
characterizes the strength and direction of the potential. For
$\alpha$ greater than zero the barrier is a ``ceiling'' and the
particle may bounce off at most once. If $\alpha$ is less than zero
then the barrier will act as a floor, and will have, in principle,
an infinite number of bounces. 
This article considers the ceiling case
for two different types of data,
\begin{itemize}
  \item Initial position, $y$, and final position, $x$;
  \item Initial momentum, $p$, and final position, $x$.
\end{itemize}
The final time is $t$, and when it is necessary to consider an 
initial time other than~$0$, it is denoted~$s$.  The time 
parameter for a path interpolating between these data is~$\tau$, 
and the corresponding position variable is~$q$; occasionally the 
notation $p(\tau)$ is needed for the momentum at some point along 
the path.

The equation governing the dynamics of the classical system and its 
general solution are
\begin{eqnarray}
\frac{\rmd^2q}{\rmd\tau ^2}=\frac{\alpha}{m}\,, \qquad
q(\tau)=\frac{\alpha}{2m}\,\tau^2+A\tau+B.\label{classic}
\end{eqnarray}
The constants of integration, $A$ and $B$, are determined from the 
initial data considered.
 For bounce trajectories two subsidiary solutions are
required, $q_1(\tau)$ and $q_2(\tau)$,
corresponding to the dynamics before and after the collision,
respectively. An extra condition must be placed on the bounce
trajectories so that the momenta of these paths at the ceiling are
equal in magnitude and opposite in direction. If $b$ is the
time at which the particle ricochets, this condition is 
\begin{eqnarray}
p_1(b)=-p_2(b).
\label{pc}\end{eqnarray}

Unless otherwise stated the remainder of the paper will use the 
natural units
\begin{eqnarray}
\hbar\equiv 1, \qquad
m\equiv {\textstyle\frac{1}{2}}\,, \qquad
\alpha\equiv 1.
\label{units}\end{eqnarray}
Note that when the last two of these equations are in force, 
position will have the
same dimensionality as time squared, and momentum will have the same
dimensionality as time:
\begin{eqnarray}
[x]=[t]^2, \qquad
[p]=[t],
\label{unitsexamples}\end{eqnarray}
 in contrast to the more familiar situation where $\hbar$ and~$c$ 
(but not~$m$) are dimensionless and hence
$[x]=[t]$, $[p] = [t]^{-1}$.
Also, the velocity is twice the momentum;
although this is classically weird, it makes the quantum formulas 
more elegant.

\subsection{The classical solutions}  \label{classsol}
\begin{figure}
\centering \caption{Three types of non-bounce trajectories with
equal initial data, $y$, and final time, $t$. The only variable is
the final position, $x$. $xd1$ represents the final position of a
type (i) trajectory; $xd2$ the final position for type (ii); and
$xt$ that for a turning trajectory, type (iii).}
\includegraphics{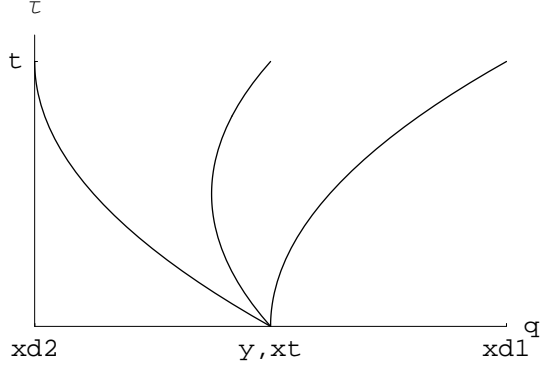}
\label{direct}
\end{figure}

The preliminary quest is to determine all possible trajectories 
connecting the
initial data $(p,s)$  or $(y,s)$  with the final data
$(x,t)$, and the corresponding constraints. For notational simplicity
 the two-point data will be abbreviated
\[
\mathbf{x}_y\equiv[(y,s),(x,t)], \qquad
\mathbf{x}_p\equiv[(p,s),(x,t)].
\]
There are three types of non-bounce trajectories to consider:
\begin{enumerate}
  \item trajectories that initially and finally move away from 
         the ceiling:
  $p\equiv p_s\in(0,\infty)$ and $p_\mathrm{f}\in(0,\infty)$;
  \item trajectories that initially and finally move toward the ceiling: 
$p\equiv p_s\in(-\infty,0)$ and $p_\mathrm{f}\in(-\infty,0)$;
  \item trajectories that initially move toward the ceiling, and end
  moving away from the ceiling: 
$p_0\in(-\infty,0)$ and $p_\mathrm{f}\in(0,\infty)$.
\end{enumerate}
A plot of these three types of  trajectory is in figure
\ref{direct}. The time at which the particle's momentum is zero,
$n$, governs the transition between motion toward the ceiling and
away from the ceiling for type (iii) trajectories.

There are certain boundary trajectories, which are
helpful in analyzing the more general situations. Trajectories of
type (i) are separated from trajectories of types (ii) and (iii)
 by a  trajectory whose initial
momentum is zero. It has initial position
\begin{equation}
\tilde{y}\equiv x-t^2,
\label{ytilde}\end{equation}
so that 
$q(\tau)=\tau^2+x-t^2$.
In this case, if
$x>t^2$ the initial position will be on the physical side of
the ceiling, but for $x<t^2$ the initial position will not
be physical. Also, keeping the final data fixed implies that if an
initial positive momentum is given to this trajectory, then the
initial position must be moved closer to the ceiling.
Conversely, for an initial negative momentum for a trajectory of
type (ii), the initial position will be pushed away from the
ceiling. Therefore, 
one  has
$y<\tilde y$ for type (i) and $y>\tilde y$ for types (ii) and (iii).

The linear potential implies that the momentum gained by the
particle is equal to the time of flight, $\tau$
(recall \eref{unitsexamples}),
 and the total
momentum of the trajectory at any time is
\begin{equation}
\tau+p.
\label{pgain}\end{equation} 
Therefore,
in case (iii) the time at which the particle will turn around is
\begin{eqnarray}
n=-p.\label{n}
\end{eqnarray}
 The distance from the initial point to the turning point
 for a trajectory where $t=n$ is
\begin{eqnarray}
\Delta=-\int_0^n v(\tau)\,\rmd\tau=
-2\int_0^{-p}(p+\tau)\,\rmd\tau = p^2=t^2.
\label{delta}
\end{eqnarray}
The negative sign occurs because the
velocity is inherently negative during the initial segment of the 
motion considered.
The boundary between cases (ii) and (iii) is marked by $y=\Delta$,
or $t=\sqrt{y}$.

\begin{figure}
\centering \caption{General bounce trajectory (dotted line)
connecting initial data, $y$, to the final data, $x$, in the same
time, $t$, as a direct trajectory (solid line).}
\includegraphics{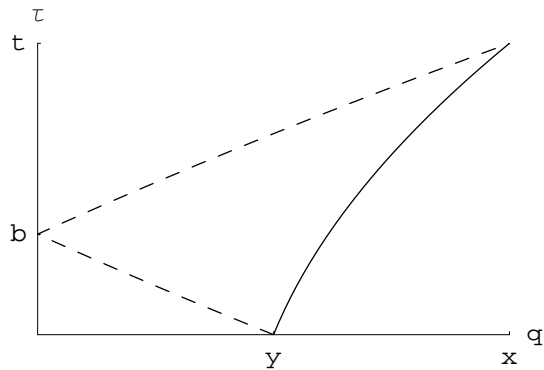}
\label{dirbou}
\end{figure}

For some initial and final data, both a direct (non-bounce) trajectory 
and
a more energetic bounce trajectory exist; see figure \ref{dirbou}.
The critical trajectories for the bounce paths are the type (iii)
trajectories with zero momentum at the ceiling, corresponding to
zero energy.
\begin{figure}
\centering \caption{Generic family of bounce trajectories with the
same initial position and final time. The final position, $x$, is
varying. The critical trajectory is dashed, and $bc$ denotes the
time of bounce for this trajectory, $xc$ denotes the final position.
The region above and to the left of the dashed curve is 
classically inaccessible.
}
\includegraphics{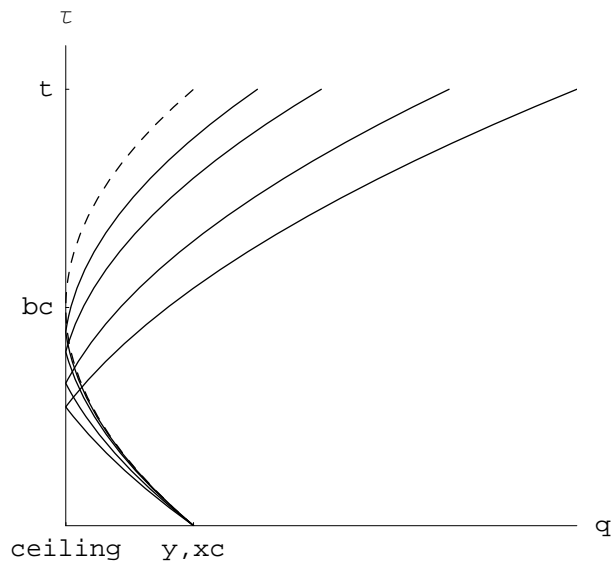}
\label{bounce}
\end{figure}
All other  trajectories for the bounce case  have energy
greater than zero, since those with negative energy can never 
reach the ceiling. 
Also, all allowed trajectories of type (iii) have negative energy, 
since otherwise they would pass through the ceiling.
 A way to view the construction of bounce trajectories is to join
two trajectories at the ceiling which there obey equation
({\ref{pc}}); one trajectory  has the correct initial data, and
the other the final data. Therefore, the critical trajectory 
 occurs when these two trajectories are the same;
the bounce trajectory becomes an ordinary path of type (iii). A particle
in a linear potential beginning at the ceiling with zero energy (and
therefore zero initial momentum) will end at location $x$ with
momentum $\sqrt{x}$. Conversely, if the particle 
arrives at the ceiling with zero energy, its momentum at the 
starting point, $y$, must have been $-\sqrt{y}$.
Therefore, by \eref{pgain} the total momentum the particle will 
gain on the critical bounce trajectory is
\begin{eqnarray}
\sqrt{x}+\sqrt{y}=t.\label{criticalbounce}
\end{eqnarray}
This equation characterizes the critical trajectory.
As will be seen in detail in \sref{position}, 
 the critical trajectory marks
the boundary between the region of space-time points $(x,t)$
 that can be connected to $y$ by classical paths and those that cannot
(see figure \ref{bounce}).

\subsection{The quantum theory}\label{eigen}

The time-dependent Schr\"odinger equation satisfied by our particle, 
in the units \eref{units}, is
\begin{equation}
-\,\pd{x^2}{^2\psi} - x\psi = \rmi \,\pd t{\psi}\,.
\label{schrod}\end{equation}
The ceiling is implemented by restricting $x$ to positive values 
and imposing the boundary condition $\psi(0,t)=0$.

The solutions of the time-independent Schr\"odinger equation,
\begin{equation}
-\,\frac{\rmd^2\psi_E}{\rmd x^2} - x\psi_E = E \psi_E\,,
\label{schrodE}\end{equation}
are Airy functions, $\Ai(-x-E)$ and $\Bi(-x-E)$ and their 
linear combinations.
The spectrum is continuous and unbounded above and below: 
$-\infty<E<\infty$.

It is known (e.g., \cite{dean_fulling})
 that the propagator that solves the initial-value problem for 
\eref{schrod} is
\begin{equation}
U(x,y,t) = \int_{-\infty}^\infty \psi_E(x)\psi_E(y) e^{-iEt} 
\rho(E) \,\rmd E
\label{ceileigen}\end{equation}
with
\begin{eqnarray}
\psi_E(x) = \pi[\Bi(-E)\Ai(-x-E) -\Ai(-E) \Bi(-x-E)], \label{eigfn}\\
 \rho(E) = \pi^{-2} [\Ai(-E)^2 + \Bi(-E)^2]^{-1}.\label{specdensity}
\end{eqnarray}
The integral \eref{ceileigen} is very oscillatory and hence hard to 
use in practice.

On the other hand, for a free particle (that is,
when \eref{schrod} holds on the whole real line, with no boundary),
the corresponding formula,
\begin{equation}
U_\infty(x,y,t) = \int_{-\infty}^\infty 
\Ai(-x-E) \Ai(-y-E) e^{-iEt}\, \rmd E,
\label{freeeigen}\end{equation}
is known in closed form (e.g., \cite{CN}, 
\cite[and references therein]{hol}):
\begin{equation}
U_\mathrm{free}(t,x,y) = (4\pi \rmi t)^{-1/2} \, 
e^{\rmi(x-y)^2/4t +\rmi(x+y)t/2 -\rmi
t^3/12}.
\label{freeprop}\end{equation}
We rederive this formula in passing in \eref{Uyd}.

For completeness, note that in the presence of a floor instead of 
a ceiling 
(that is, $\alpha=-1$),
the spectrum is discrete and the propagator is easily seen to take 
the form
\begin{equation}
U_\mathrm{floor}(x,y,t) = \sum_{n=1}^\infty 
\Ai'(-E_n)^{-2} \Ai(x-E_n)\Ai(y-E_n) e^{-iE_n t},
\label{flooreigen}\end{equation}
where the $E_n$ are the roots of
$ \Ai(-E_n)=0$. 
The WKB solution in this case would be difficult, because of the 
need to sum over many reflected  classical paths 
(as the particle bounces inside the potential well).

\subsection{The WKB ansatz}

In this section we reintroduce the constants $\hbar$ and $m$ and 
allow for a 
higher-dimensional configuration space;
thus $x$ becomes $\mathbf{x}$, and we usually omit the arguments 
$y$, $s$, and~$t$.
The mathematical correspondence between classical physics and the
quantum propagator begins with the ansatz
\begin{equation}
U(\mathbf{x})=
\e^{\frac{\rmi}{\hbar}S(\mathbf{x})}\sum_{j=0}^\infty
\left(\frac{\rmi}{\hbar}\right)^{-j}A_j(\mathbf{x})\label{wkbansatz}
\end{equation}
for a solution to the time-dependent Schr\"{o}dinger equation
\begin{equation}
-\,\frac{\hbar^2}{2m}\nabla^2\psi+V(x)\psi=\rmi\hbar\frac{\partial
\psi}{\partial t}\,.\label{schrodinger}
\end{equation}

The resulting $\emph{O}(\hbar^0)$ equation is the Hamilton--Jacobi
equation of classical physics,
\begin{equation}
\frac{1}{2m}\left(\nabla S(\textbf{x})\right)^2+V(x)=
-\,\frac{\partial S(\textbf{x})}{\partial
t},\label{wkbphase}
\end{equation}
whose solution, the action $S(\mathbf{x})$,  is the phase of the 
WKB propagator and also the
generating function for the Lagrangian manifold corresponding to the
classical system \cite{litjohn}:
\begin{equation}
\mathbf{p}(\mathbf{x})\equiv\mathbf{p}(t,\mathbf{x})
=\nabla S(\mathbf{x}),  \qquad
\mathbf{p}(0,\mathbf{y}) = -\nabla_{\mathbf{y}}S.\label{lman_p}
\end{equation}

The amplitude of the propagator is the solution to the
${\emph{O}}(\hbar)$ equation resulting from (\ref{wkbansatz}),
\begin{equation}
A(\textbf{x})\nabla^2 S(\textbf{x}) 
+2 \nabla S(\textbf{x}) \cdot \nabla A (\textbf{x}) 
=-\,\frac{\partial A(\textbf{x})}{\partial t}\,.\label{wkbamp}
\end{equation}
The quantity $\rho({\textbf{x}})\equiv\abs{A(\textbf{x})}^2$
is interpreted as the probability density associated with an
ensemble of classical particles. The
initial amplitude, $\rho_0(\textbf{x})$, (i.e., the amplitude
corresponding to $t\to s=0$, $\textbf{x}\to\textbf{y})$
 should agree with the correct
initial form of the quantum propagator, and the evolution of
$\rho(\textbf{x})$ obeys
\begin{eqnarray}
m\,\pd{t}{\rho\left(\mathbf{x}\right)}=-\nabla\cdot
\left[\rho\left(\mathbf{x}\right)\,
\mathbf{p}\left(\mathbf{x}\right)\right]\,,\label{continuity_eq}
\end{eqnarray}
where the first of equations (\ref{lman_p}) defines the momentum
field, $\mathbf{p}\left(\mathbf{x}\right)$.
Equation (\ref{continuity_eq}) is a continuity equation for the
probability density function,
and it can be shown that the evolution of the density function
 from the initial configuration, $\mathbf{x}_0\equiv\mathbf{y}$,
  to the final configuration, $\mathbf{x}$, is
\begin{eqnarray}
\rho_0(\textbf{x}_0)\,\rmd \mathbf{x}_0=\rho(\textbf{x})\,\rmd \mathbf{x} 
 \;\Rightarrow\;  
\rho(\textbf{x})=\rho_o(\textbf{x}_0)
\abs{\frac{\partial \mathbf{x}_0}{\partial
\mathbf{x}}}\,.\label{amptra}
\end{eqnarray}
In the 1-dimensional case the Jacobian
 $\frac{\partial \mathbf{x}_o}{\partial\mathbf{x}}$ becomes $\pd{x}{y}$.

For the propagator based on trajectories with fixed initial position $y$,
 equation (\ref{amptra}) becomes inapplicable; 
the trajectories are labeled by their initial 
momenta, $p\equiv p(0)\equiv p(0,x)$, and the Jacobian must be 
replaced by 
\begin{eqnarray}\left|\frac{\partial p}{\partial x}\right| = 
\left| \frac{\partial^2 S}{\partial x \,\partial y}\right| =
\left| \frac{\partial p(t)}{\partial y}\right|,
\label{wkb_amp}
\end{eqnarray}
where \eref{lman_p} has been used.
In that situation the second of equations \eref{lman_p} is replaced by
\begin{equation}
y = \frac{\partial S}{\partial p}\,.
\label{Sreversed} \end{equation}

This paper will not consider solutions of (\ref{wkbansatz}) higher
than $O(\hbar)$. 
The WKB approximation for the propagator to first order in $\hbar$ is
therefore
\begin{equation}
U(\textbf{x})=A(\textbf{x})\,\rme^{\frac{\rmi}{\hbar}S(\mathbf{x})}\,,
\end{equation}
and is subject to the same initial conditions as the corresponding
quantum propagator. 
For trajectories with initial position data the initial value is
(in dimension~$1$)
\begin{eqnarray}
\lim_{t\rightarrow s}U(x)=\delta(x-y)\,,
\end{eqnarray}
and for trajectories described with initial momentum data the 
initial value of the propagator is
\begin{eqnarray}
\lim_{t\rightarrow s}U(p)=\frac{\rme^{\frac{\rmi}{\hbar}xp}}
{\sqrt{2\pi\hbar}}\,.
\end{eqnarray} 

It is often said that the WKB approximation is exact whenever the 
Hamiltonian is a polynomial of degree at most $2$ in both 
position and momentum.  That statement is true, however, only for 
the propagator, not for solutions of the Schr\"odinger equation 
with more general initial data \cite{BS}, and only when the 
Hamiltonian is globally of such a form.  In our problem we shall 
find in \sref{position} that the WKB solution is exact until the 
wave hits the ceiling but has
 only $O(\hbar)$ accuracy for the reflected wave.

\section{The propagator for  initial position data}\label{position}
Throughout this section 
\begin{eqnarray}
\mathbf{d}_y\equiv\left[(y,0),(x,t)\right], \quad
\mathbf{b}_y\equiv\left[(y,0),(0,b_y)\right], \quad
\mathbf{r}_y\equiv\left[(0,b_y),(x,t)\right].
\label{xdatapoints}\end{eqnarray}
Here $b_y$ denotes the time of bounce for a trajectory starting at~$y$.
In our considerations $x$, $y$, and $t$ are all nonnegative.

The general trajectory and momentum connecting the initial data
$(y,s)$ to the final data $(x,t)$ for a
linear potential are
\numparts\begin{eqnarray}
q(\tau;\mathbf{x}_y)=
\left(\tau-s\right)^2+\left(\tau-s\right)
\left[\frac{x-y}{t-s}-\left(t-s\right)\right]+y,
\label{generalqy}\\
p(\tau;\mathbf{x}_y)=
\left(\tau-s\right)+\frac{1}{2}\left[\frac{x-y}{t-s}
-\left(t-s\right)\right].
\label{generalpy}
\end{eqnarray}
\endnumparts
For the direct paths, $s=0$ in $\mathbf{x}_y\,$, and the
trajectory and momentum are
\begin{eqnarray}
q(\tau;\mathbf{d}_y)=
\tau^2+\tau\left[\frac{x-y}{t}-t\right]+y\label{specificqy},\\
p(\tau;\mathbf{d}_y)=
\tau+\frac{1}{2}\left(\frac{x-y}{t}-t\right).\label{specificpy}
\end{eqnarray}
Note that equation (\ref{specificpy}) implies that a trajectory
with initial momentum $0$ will have initial position $\tilde y$,
 as predicted \eref{ytilde}.
Using equations \eref{n}, \eref{specificpy}, and \eref{specificqy},
 one sees that the time at which the particle  turns around, 
$n(x,y,t)$, is
\begin{eqnarray}
n=-p=-\left(\frac{x-y}{2t}-\frac{t}{2}\right) =
\frac{1}{2}\left(\frac{y-x}{t}+t\right),
\label{n1}\end{eqnarray}
and the place where this event occurs is
\begin{equation}
q(n) = -\,\frac14\left(\frac{x-y}t -t\right)^2 +y.
\label{qb}\end{equation}

\subsection{Admissible paths}
In our system the path \eref{specificqy}  is admissible if and 
only if it does not penetrate the ceiling during the time interval 
$(0,t)$.
If $n\le0$, the path is admissible and of type~(i);
if $n\ge t$, it is admissible and of type~(ii).
If $0<n<t$ and $q(n)\ge0$, the path is admissible and of type (iii); 
if
\begin{equation}
 0 < n < t\quad\textrm{and}\quad q(n) < 0 ,
\label{bouncecond}\end{equation}
it is forbidden.
After some calculation these two conditions  respectively translate to
\begin{equation}
t^2 > |x-y|,
\label{bbound}\end{equation}
\begin{equation}
(x+y-t^2)^2 > 4xy. 
\label{qbound}\end{equation}

The solution of \eref{qbound} falls naturally into two cases:
If $t^2>x+y$, then \eref{qbound} is equivalent to 
$t > \sqrt{x} +\sqrt{y}$, 
and \eref{bbound} is then satisfied.
Conversely, if $t \le \sqrt{x} +\sqrt{y}$ and $t^2>x+y$, then $q(n)\ge 0$ 
and the trajectory is allowed.
If $t^2 \le x+y$, \eref{qbound} is equivalent to 
$t^2\le |\sqrt{x}-\sqrt{y}|^2$, 
which is easily shown to contradict \eref{bbound}; 
so all trajectories are allowed in this case.

In summary, all allowed trajectories satisfy 
\begin{equation}
t\le \sqrt{x}+\sqrt{y}
\label{tbound}\end{equation}
and all forbidden trajectories violate it, 
in accordance with the energy argument at the end of \sref{classsol}.
In the next subsection we verify that \eref{tbound} is also the necessary 
and sufficient condition for existence of a bounce trajectory.
For given data $(y,x,t)$, therefore, there will be either one direct 
path and one bounce path, or no path at all.  The only exception 
is the critical case \eref{criticalbounce}, for which the direct and 
bounce paths are the same.

A finer classification of paths is given in Table \ref{tablex},
which is useful for comparison with the listing of paths of 
prescribed initial momentum in Table \ref{tablep} in 
\sref{momentum}.  Our tables do not include paths 
corresponding to endpoints of the parameter intervals listed, 
partly because the classification of those can be ambivalent.
The classification is more symmetrical in $x$ and $y$ than it may 
appear:
Interchanging $x$ with $y$ interchanges type~(i) with 
type~(ii), and it leaves the type-(iii) region, as a whole, 
invariant, its boundaries being the lines $t^2=|x-y|$ and
a segment of the parabola $(x-y)^2 -2(x+y)t^2+ t^4=0$.

\Table{\label{tablex} Constraints on the initial position given 
the 
final data $\left(x,t\right)$. Endpoint cases are not included.}
\br
$x\in$
&$y\in$&Trajectory\\
\mr
$\left(t^2,\infty\right)$&$\left(0,x-t^2\right)$&Type (i) 
(rightward)\\
$\left(0,\infty\right)$&$\left(x+t^2,\infty\right)$&Type (ii)
(leftward)\\
$\left(0,t^2\right)$&$\left(\left(\sqrt{x}-t\right)^2,x+t^2\right)$
&Type (iii) (turning)\\
$\left(t^2,\infty\right)$&$\left(x-t^2,x+t^2\right)$&Type (iii)
(turning)\\
$\left(0,\infty\right)$&$\left(\left(t-\sqrt{x}\right)^2,\infty\right)$
&Bounce
\\\br
\endTable 

\subsection{The time of bounce}  \label{bouncetime} 
   From equations (\ref{generalpy}) and (\ref{generalqy}) the
trajectories and momenta for the bounce paths are
\numparts
\begin{eqnarray}
q_1(\tau;\mathbf{b}_y)=\tau^2-\tau\left(
b+\frac{y}{b}\right)+y, \label{5}\\
q_2(\tau;\mathbf{r}_y)=\left(\tau^2-t^2\right)+\frac{\tau-t}{b-t}
\left(-\left(b^2-t^2\right)-x\right)+x, \label{6}\\
p_1(\tau;\mathbf{b}_y)=\tau-\left(\frac{y}{2b}+\frac{b}{2}\right), 
\label{6a}\\
p_2(\tau;\mathbf{r}_y)=\tau-\frac{1}{2}\left(\frac{x}{b-t}
+\left(b+t\right)\right).
\label{6b}
\end{eqnarray}
\endnumparts
Thus the  ceiling condition for the bounce trajectory,
$q_{1}(b)=-q_{2}(b)$, yields the
equation
\begin{equation}
f(b)\equiv b^3+a_2b^2+a_1b+a_0=0,
\label{cubic1}\end{equation}
\[ 
a_2\equiv-\,\frac{3t}{2}\,, \quad
a_1\equiv\frac{t^2}{2}-\frac{1}{2}\left(x+y\right), \quad
a_0\equiv\frac{yt}{2}\,. 
\]

The polynomial discriminant of the cubic equation, $D$, is defined
as \cite{Cubic}
\begin{eqnarray} 
D&\equiv R^2+Q^3 \label{Discriminant} \\ 
&=
\frac{t^2}{64}\left((x-y)^2-\frac{4}{9}(x+y)^2\right)
-\left(\frac{t^2}{12}\right)^3-\left(\frac{1}{6}(x+y)\right)^3
-\frac{t^4}{288}(x+y),  \nonumber
\end{eqnarray}
where the definitions of $R$ and $Q$ are
\numparts
\begin{eqnarray}
Q&=&\frac{3a_1-a_2^2}{9}=-\frac{t^2}{12}-\frac{1}{6}(x+y)\label{Q},\\
R&=&\frac{9a_1a_2-27a_0-2a_2^3}{54}=\frac{t}{8}(x-y)\label{R}.
\end{eqnarray}
\endnumparts
If $D>0$ then one root is real and the other two are complex
conjugates; $D=0$ if all roots are real and at least two are equal;
and $D<0$ if all roots are real and unequal. From equation
(\ref{Discriminant}),
\begin{eqnarray}
D(0)<0
\qquad\textrm{and}\qquad
\lim_{t\rightarrow\infty}D(t)=-\infty. \label{dis}
\end{eqnarray}
To analyze the discriminant's behavior as $t\rightarrow \infty$ let
$T\equiv t^2$; then
\begin{eqnarray}
\frac{dD}{dT}=\frac{1}{64}\left(\left(x-y\right)^2-\frac{4}{9}
\left(x+y\right)^2\right)
-\left(\frac{T}{24}\right)^2-\frac{T}{144}\left(x+y\right).\label{dDdt}
\end{eqnarray}
Therefore, the derivative of the discriminant will tend to $-\infty$
as $t\rightarrow\infty$. Setting \eref{dDdt} to zero reveals
that the critical points of the discriminant occur where
\begin{equation}
t = \pm\sqrt{\left(x(-2\pm3)-y(2\pm3)\right)}. 
\label{discrderi} \end{equation}
The two negative roots are not physical, so the only possible
critical points are
\begin{eqnarray}
t_{\mathrm{c}+}=\sqrt{x-5y}, \qquad
t_{\mathrm{c}-}=\sqrt{y-5x}.
\label{critpts}\end{eqnarray}
If the initial and final data do not satisfy $x>5y$ or $y>5x$, then
the derivative of the discriminant will always be negative, and by
\eref{dis}  so will the discriminant. If one of these inequalities
is satisfied, 
 then only
one of the  roots \eref{critpts} will be real. 
Without loss of generality, assume $x>5y$,
so that the relevant root of is $t_{\mathrm{c}+}$.
Furthermore, relations \eref{dis} imply that this
root must be a maximum for $D(t)$. 
Therefore, if $D(t_{\mathrm{c}+})<0$, then
the discriminant will be negative for all values of~$t$.
According to (\ref{Discriminant}),
\[
D(t_{\mathrm{c}+})=\frac{1}{64}\left(\left(x-5y\right)
\left(x-y\right)^2-\left(x-y\right)^3\right). 
\]
Thus the discriminant will always be negative if
$x-5y<x-y$, or $y>0$.
Therefore, provided that $y\neq0$, $D(t)<0$ for all $t$  and all
three roots to equation (\ref{cubic1}) are real and unequal.
Similarly, if $y>5x$ then the requirement that $D(t)<0$ is $x>0$,
which is in general true except for the special case $x=0$. The
special cases $y=0$ and $x=0$ respectively imply that the initial
and final positions are at the ceiling; the respective times of bounce
are $0$ and~$t$.

The maximum and minimum of the cubic function \eref{cubic1} are
\[
b_{\mathrm{c}\pm}=\frac{t}{2}\pm\sqrt{\frac{t^2}{12}+\frac{1}{6}(x+y)}\,,
\]
and the point at which the equation changes its concavity is
$\frac{t}{2}\,$. Since $b\rightarrow\pm\infty$ implies
$f(b)\rightarrow\pm\infty$, the point $b_\mathrm{c}$ is where $f(b)$ 
changes
from concave to convex. Since all three roots must be real, one
root of $f(b)$ will lie between the two extrema, and the other two
must lie outside of the range:
\begin{eqnarray}
r_1\in\left(-\infty,b_{\mathrm{c}-}\right), \qquad
r_2\in\left(b_{\mathrm{c}+},\infty\right), \qquad
r_3\in\left(b_{\mathrm{c}-},b_{\mathrm{c}+}\right).
\label{r3}\end{eqnarray}
Since $r_3$ is the only root which may equal $t$ and $0$, which are
the critical values of the bounce time, and the bounce time is a
continuous function, $r_3$ is the correct root for all trajectories.
Therefore, if a root of $f(b)$ is found which agrees with the
critical  trajectories, 
it will be the correct root for all the trajectories.

For the case where $D<0$ the three cubic roots may be written as
(e.g., \cite{Cubic,Namias})
\begin{equation}
r_j\left(x,y,t\right)=\frac{t}{2}+2\sqrt{-Q(x,y,t)}
\cos\left(\frac{\Theta\left(x,y,t\right)+2j\pi}{3}\right),  
\label{root}
\end{equation}
where  $j=0,1,2$ and 
\begin{eqnarray}
\Theta(x,y,t)=\arccos\left(\frac{R(x,y,t)}{\sqrt{-Q(x,y,t)^3}}\right).
\end{eqnarray}
Finally, for a trajectory in which $x=y$, the correct bounce time is
$b_y=\frac{t}{2}\,$, and only the last range in (\ref{root}) contains
this value. Therefore, by the considerations above
\begin{eqnarray}
b_y &= r_2(x,y,t)   \nonumber \\
&= \frac t2 + \frac1{\sqrt{3}}\sqrt{t^2 + 2(x+y)} 
\sin\left[\frac13\,\sin^{-1}
{3\sqrt3 \,t(y-x) \over (t^2+ 2(x+y))^{3/2}} \right].
\label{sinroot}\end{eqnarray}

\subsection{The action}\label{xyaction}
To complete the WKB construction the action of the trajectories and
the amplitude function must be determined. From equations
(\ref{generalqy}) and (\ref{generalpy}) the general Lagrangian for
the initial-position formulation is
\begin{eqnarray}
\fl L\left[q,\dot{q},\tau\right]&=p(\tau;\mathbf{x}_y)^2+
q(\tau;\mathbf{x}_y)\nonumber\\
\fl
&=2\left(\tau-t_0\right)^2+2\left(\tau-t_0\right)
\left[\frac{x-y}{t-t_0}-\left(t-t_0\right)\right]+
\frac{1}{4}\left[\frac{x-y}{t-t_0}-\left(t-t_0\right)\right]^2
+y\label{xyLgen},
\end{eqnarray}
and the corresponding action is
\begin{eqnarray}
S_y(t;\mathbf{x}_y)&=&
\frac{2}{3}\left(t-t_0\right)^3+\left(t-t_0\right)^2
\left[\frac{x-y}{t-t_0}-\left(t-t_0\right)\right]\nonumber\\
&&{}+\left(t-t_0\right)\left(\frac{1}{4}
\left[\frac{x-y}{t-t_0}-\left(t-t_0\right)\right]^2+y\right).
\label{xySgen2}
\end{eqnarray}
Therefore, the actions for the direct and bounce trajectories are
\numparts
\begin{eqnarray}
S_{\mathbf{d}y}=S_y(t;\mathbf{d}_y),\label{xySd}\\
S_{\mathbf{b}y}=S_y(b_y;\mathbf{b}_y)+S_y(t;\mathbf{r}_y).\label{xySb}
\end{eqnarray}
\endnumparts
Note that for the direct case, as $t\rightarrow0$ the action becomes
\begin{eqnarray}
\lim_{t\rightarrow0}S_y(t;\mathbf{d}_y)\rightarrow\
frac{\left(x-y\right)^2}{4t}\,,
\end{eqnarray}
which is the free particle action.

\subsection{The amplitude}
   From equations \eref{specificpy} and \eref{wkb_amp},
 the amplitude for the direct trajectory is 
\begin{eqnarray}
\sqrt{\frac{1}{2\pi\rmi}}\sqrt{\frac{\partial p(t;\mathbf{d}_y)}
{\partial y}}
=\frac{1}{\sqrt{4\pi\rmi t}}\,.
\label{directamp}\end{eqnarray}
Note that as  $(x,t)\rightarrow (x,0)$ the constructed amplitude 
yields the correct initial condition, $\delta\left(y-x\right)$:
\begin{eqnarray}
\lim_{t\rightarrow
0}\frac{1}{\sqrt{4\pi\rmi t}}\,e^{\rmi S_d\left(x,y,t\right)}=
\lim_{t\rightarrow
0}\frac{1}{\sqrt{4\pi\rmi t}}e^{\frac{\left(y-x\right)^2}{4\rmi t}}
=\delta\left(y-x\right)\,.
\nonumber\end{eqnarray}

Using $r_2$ from equation (\ref{root}) and $p_1\left(0\right)$ from
equation (\ref{6a}), we find the Jacobian corresponding to the action for
the bounce case:
\begin{eqnarray}
\fl\frac{\partial p(0;\mathbf{b}_y)}{\partial
x}&=&\left(\frac{y}{2b_y^2}-\frac{1}{2}\right)\frac{\partial
b_y}{\partial x}\nonumber\\
\fl&=&\left(\frac{y}{2b_y^2}-\frac{1}{2}\right)
\frac{\cos\left(\frac{\Theta+4\pi}{3}\right)}
{6\sqrt{-Q}}\left[1-\frac{Q}{6\sqrt{-D}}
\left(\frac{t}{2}+\frac{R}{Q^4}\right)
\tan\left(\frac{\Theta+4\pi}{3}\right)\right].
\end{eqnarray}
However, linearizing the classical equations of motion for the
potential $V=-\abs{q}$, and then ``folding'' the negative half of
the plane onto the positive half produces a more explicit form of
the Jacobian. For the given potential we have
\begin{eqnarray}
-\frac{\partial V}{\partial
q}=\sgn(q)=2\theta(q)-1,\label{jacobi3}
\end{eqnarray}
where $\theta(q)$ is the Heavyside step function.
Differentiating (\ref{jacobi3}) with respect to a parameter,
$\alpha$, yields
\begin{eqnarray}
\frac{\partial}{\partial\alpha}\left[2\theta(q)-1\right]=2\delta(q)=
2\,\frac{\delta(\tau-b_y)}{\abs{\dot{q}(b_y)}}\,\frac{\partial
q}{\partial \alpha}\,.
\end{eqnarray}
where we have used the functional dependence of the trajectory
 $q(\tau)$ and the fact that
$q(b_y)=0$.  Using equation (\ref{6a}) and
the fact that $p(b_y)<0$ for the trajectory
moving toward the ceiling, we have
\begin{eqnarray}
\abs{\dot{q}(b_y)}=\frac{y}{b_y}-b_y\,.\label{jacobi4}
\end{eqnarray}
Therefore, differentiating Hamilton's equations with respect
$\alpha\equiv y$ yields
\numparts
\begin{eqnarray}
\frac{\rmd}{\rmd\tau}\frac{\partial p}{\partial
 y}=2\,\frac{\delta(\tau-b_y)}{\abs{\dot{q}(b_y)}}\frac{\partial
q}{\partial  y}\,,\label{jacobi1}\\
\frac{\rmd}{\rmd\tau}\frac{\partial q}{\partial y}=2\frac{\partial
p}{\partial y}\,.\label{jacobi2}
\end{eqnarray}
\endnumparts
For the trajectory moving toward the ceiling, $\tau<b_y\,$, equation
(\ref{jacobi1}) predicts
\begin{eqnarray}
\frac{\rmd}{\rmd\tau}\frac{\partial p}{\partial y}=0,\nonumber
\end{eqnarray}
so that $\frac{\partial p}{\partial y}$ is a constant. Therefore,
\begin{eqnarray}
\frac{\partial p}{\partial y}(\tau)=C \equiv 
\frac{\partial p}{\partial y}(0)
\end{eqnarray}
for $\tau<b_y\,$, and hence the solution to (\ref{jacobi3}) for
$\tau<b_y$ is
\begin{eqnarray}
\frac{\partial q}{\partial y}=2C\tau+1,
\end{eqnarray}
since $q(0)=y$ implies $\frac{\partial q}{\partial y}(0)=1$.
Integrating (\ref{jacobi1}) for $\tau>b_y$ and using equation
(\ref{jacobi4}) yields
\begin{eqnarray}
\frac{\partial p}{\partial
y}=C+\frac{2}{\abs{\dot{q}(b_y)}}\frac{\partial
q}{\partial y}(b_y) 
=\frac{C\left[y+3b_y^2\right]+2b_y}{y-b_y^2}\,.\label{dpdy>b}
\end{eqnarray}
And using this result for $\frac{\partial q}{\partial y}$ yields
\begin{eqnarray}
\frac{\partial q}{\partial y}(\tau)=\frac{\partial q}{\partial
y}(b_y)+2\frac{\partial p}{\partial
y}\int_{b_y}^t d\tau
=\frac{\partial q}{\partial y}(b_y)+2\frac{\partial
p}{\partial y}\left(\tau-b_y\right).\nonumber
\end{eqnarray}
Since $q(t)=x$ it follows that $\frac{\partial q}{\partial y}(t)=0$.
Therefore, after algebraic manipulations, the constant
$C$ is
\begin{eqnarray}
\frac{\partial p}{\partial y}(0)=
\frac{1}{2}\,\frac{5b_y^2-4tb_y-y}{\left(y-b_y^2\right)b
+(t-b_y)\left(y+3b_y^2\right)}\,.
\end{eqnarray}
Finally, putting this result into (\ref{dpdy>b}) and using the 
cubic equation
to simplify the denominator, we get the desired Jacobian:
\begin{eqnarray}
-\frac{\partial p}{\partial
y}(t)=\frac{b_y^2-y}{2\left[-3tb_y^2+2(t^2-x-y)b_y+3yt\right]}\,;
\label{dpdyt}\end{eqnarray}
the minus sign comes from ``folding'' the negative half-plane over.

The initial amplitude function for the bounce trajectories must match
in magnitude the one  for the direct trajectories, \eref{directamp}.
Therefore, the final form of the
amplitude function for bounce paths is, up to a possible phase,
\begin{eqnarray}
A_{yb}=\frac{1}{\sqrt{2\pi\rmi}}
\abs{\frac{b_y^2-y}{2\left[-3tb_y^2+2(t^2-x-y)b_y+3yt\right]}}^
{\frac{1}{2}}.
\label{Aby}\end{eqnarray}

Equation (\ref{criticalbounce}) predicts that on the critical
curve, $x=t^2+y-2t\sqrt{y}$. Since the critical curve for the bounce
path is identical with the trajectory of type (iii) for $E=0$, the
time of bounce is given by $n_y$ on the critical curve. And
substituting for $x$ into equation (\ref{n1}) yields
\begin{eqnarray}
y=b_y^2
\end{eqnarray}
on the critical curve. Therefore, the amplitude for the bounce path
will vanish on the critical curve.
Below the critical curve (in the ``allowed'' region) $y>b_y^2$,
so the numerator in \eref{Aby} is negative.

On the ceiling we want the bounce propagator to agree numerically
(up to sign, see below) with the direct propagator of
which it is a continuation.  This limit corresponds to
$x=0$ and $b_y=t$.  One can check (see \eref{Uyd} and \eref{Uyb} below)
that the actions agree there, and that the denominator in the bounce 
amplitude reduces as
\[
-3tb_y^2+2(t^2-x-y)b_y +3yt
=-3t^3+2t^4-2yt+3yt = t(y-b_y^2),
\]
so that the amplitudes match as well.

Putting everything together we have the WKB propagators for the direct 
and bounce cases:
\numparts
\begin{eqnarray}
\fl U_{y\mathrm{d}}(x,y,t)&=\frac{1}{\sqrt{4\pi \rmi t}}
\exp\left[\rmi\left(\frac{2}{3}t^3+t^2\left[\frac{x-y}{t}-t\right]+
t\left(\frac{1}{4}\left[\frac{x-y}{t}-t\right]^2+y\right)\right)\right],
\label{Uyd}\\
\fl U_{y\mathrm{b}}\left(x,y,t\right)&=\frac{1}{\sqrt{4\pi\rmi}}
\left[{\frac{y-b_y^2}{-3tb_y^2+2(t^2-x-y)b_y+3yt}}\right]^
{\frac{1}{2}}\nonumber\\
\fl&\times\exp\left[\rmi\left(\frac{2}{3}b_y^3
-b_y^2\left[\frac{y}{t}+b\right]+
b\left(\frac{1}{4}\left[\frac{y}{t}+b_y\right]^2
+y\right)\right)\right]\nonumber\\
\fl&\times\exp\left[\rmi\left(\frac{2}{3}(t-b_y)^3+(t-b_y)^2
\left[\frac{x}{t}-(t-b_y)\right]\right)\right]\nonumber\\
\fl&\times\exp\left[\rmi\,\frac{(t-b_y)}{4}\left(\frac{x}{t}
-(t-b_y)\right)^2\right].
\label{Uyb}\end{eqnarray}
\endnumparts
Note that, with the exception of the constraints on the initial
position, the propagator for the direct paths is identical with that
for the linear potential without a ceiling, as given in \eref{freeprop}. 
The complete propagator obeying the Dirichlet boundary condition,
$\psi(0,t)=0$, must vanish at the ceiling.  Therefore, its correct WKB
approximation is the \emph{difference},
$U_{y\mathrm{d}}-U_{y\mathrm{b}}\,$, if the phase convention for the
bounce propagator is that adopted in \eref{Uyb}.
The \emph{sum} solves, to lowest order in $\hbar$, the Neumann problem,
$\frac{\partial\psi(0,t)}{\partial x} =0$
(by virtue of \eref{lman_p} and \eref{pc}).
(In more technical language, \eref{Uyb} does not include the Maslov index,
or, rather, its analogue for a sharp boundary.)

\section{Soft ceilings} \label{caustic}

What is happening at the critical curve, $t = \sqrt{x}+\sqrt{y}$, 
is made clearer by studying a smoother model.  We replace the hard 
ceiling (Dirichlet boundary condition) by a smooth but steeply 
rising potential. For algebraic convenience we place the ceiling at 
$x=1$ and the barrier on the right instead of the left; 
then the potential function
\begin{equation}
V(x) = x + x^n
\label{softpot}\end{equation}
for some large $n$ does what we want.  
We take $n$ to be an even integer.  Then
as $n\to+\infty$, the term $x^n$ vanishes for $|x|<1$ and approaches 
infinity for $|x|>1$.
The barrier at $x=-1$ was not present in our original scenario 
(where it would have been at $x=2$), 
but it will not affect the classical solutions in the regime where 
we shall examine them.

\begin{figure}
\centering\caption{A family of paths leaving the origin with various 
initial momenta, for $n=6$ in \eref{softpot}.
Formation of a fold caustic at the top of the figure is clearly visible.
Horizontal axis is $t$, vertical is $x$, $y$ is fixed.}
\includegraphics{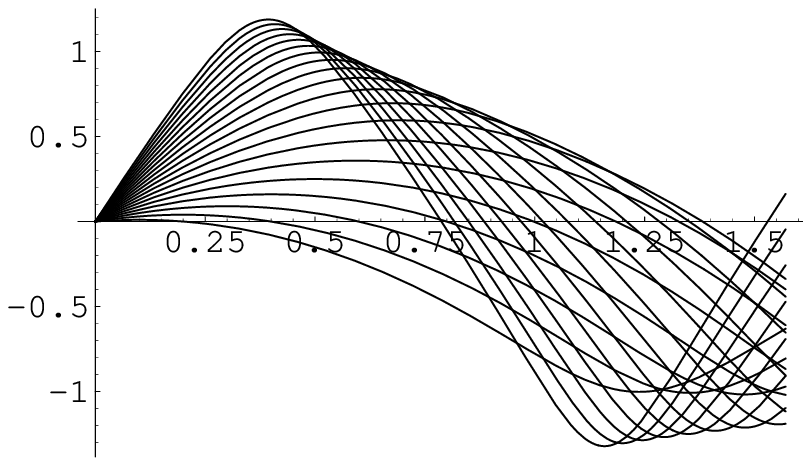} \qquad \includegraphics{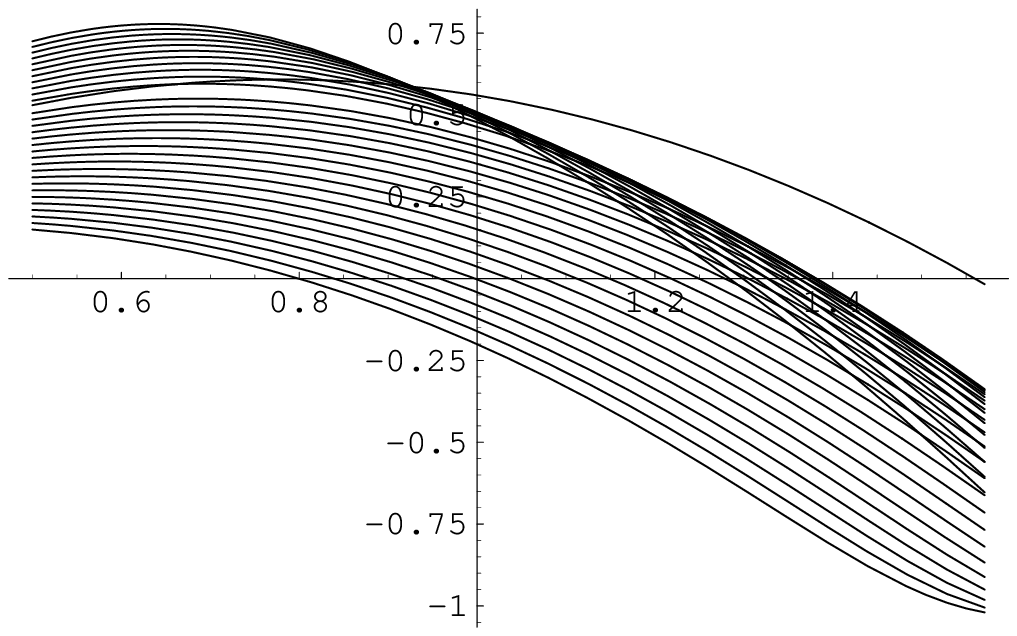}
\label{pot6}\end{figure}

\begin{figure}
\centering\caption{The same for $n=30$.  
The leftmost part of the caustic curve is converging to  a ``ceiling'' 
at $x=1$.  
The trajectory in the upper right is close to the critical trajectory 
\eref{criticalbounce} and can be loosely described as 
``the last path that misses the caustic''\negthinspace.}
\includegraphics{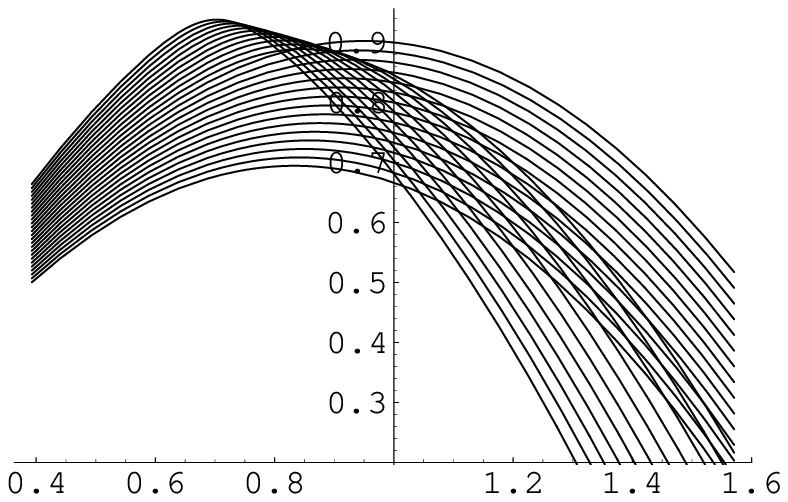} \qquad 
\includegraphics{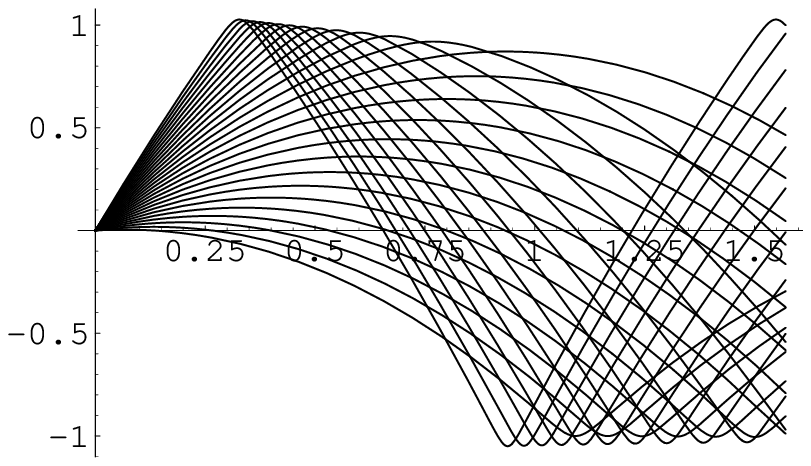}
\label{pot30}\end{figure}

A potential of this type creates a conventional caustic to which 
the Maslov theory applies. As $n\to\infty$, part of the caustic 
curve converges to the critical curve while the rest of the 
caustic converges to the initial portion of the ceiling.  The 
families of classical trajectories starting from $y=0$ are 
displayed by {\sl Mathematica\/} in Figures \ref{pot6} and 
\ref{pot30} for $n = 6$ and $30$, respectively.  The stray curves 
in the upper right of the closeup plots, \ref{pot6}(b) and 
\ref{pot30}(b), are artifacts of instability in the numerical 
solution of the differential equation; it is not surprising that 
these occurred very close to the caustic limit where various 
solutions are nearly tangent. (The physical invalidity of these 
curves is clear from the observation that they intersect other 
trajectories that started with greater kinetic energy but now 
appear to have less.) The returning trajectories in the lower 
right of plots \ref{pot6}(a) and \ref{pot30}(a), on the other 
hand, are artifacts of the model, representing reflection off the 
gratuitous floor at $x=-1$ introduced by the potential 
\eref{softpot}.

Disregarding these extraneous features, one clearly sees a 
standard fold caustic, the envelope of the trajectories, 
developing at the top of the figures. Below the caustic each 
point has two trajectories through it, which we can label 
``bounce'' or ``direct'' according to whether or not the path is 
returning from touching the caustic. For a fixed $y$, all 
sufficiently energetic paths appear, for large~$n$, to bounce off 
the horizontal line $x=1$, which becomes the ceiling. Less 
energetic paths do not reach all the way to the ceiling, but on 
their return they cross over paths of still lower energy, 
creating an envelope close to the critical trajectory of the 
hard-ceiling problem.

If we were to construct the semiclassical propagator for this 
system, we would presumably find that it is a poor approximation 
near the caustic, in the sense that when substituted into the 
time-dependent Schr\"odinger equation \eref{schrod}) it leaves a 
large residual. At points well inside the caustic ($\sqrt{x} \ll 
t - \sqrt{y}$) it should be a good approximation if direct and 
bounce contributions are combined with the proper phase. 
Nevertheless, as $n\to\infty$ the contribution of the direct 
paths alone will converge to $U_{y\mathrm{d}}$ in \eref{Uyd}, 
which we have seen to be an exact solution of \eref{schrod} 
(though not of the ceiling boundary condition).  That is, the 
region where $U_{y\mathrm{d}}$ is a poor solution is pressed into 
the caustic boundary as $n\to\infty$, as \eref{softpot} 
increasingly well approximates a pure linear potential at 
smaller~$x$.

Near a typical caustic the amplitude function \eref{wkb_amp} of 
the semiclassical approximation diverges, since it is the 
Jacobian determinant of the mapping from initial momentum to 
final position (which is by definition singular on the caustic).  
As just remarked, this is not true of the semiclassical solution 
for the hard wall, \eref{Uyd} and \eref{Uyb}. The amplitude of 
\eref{Uyb} actually vanishes on the critical curve.  It does, 
however, have a square-root singularity there, so that its 
residual in \eref{schrod} (which involves second derivatives of 
the amplitude) will still blow up.  For this reason, and because 
the position-space semiclassical construction yields no nonzero 
prediction at all for the region of space-time beyond the 
caustic, we turn in the next section to a momentum-space 
construction.

\section{The propagator for initial momentum data}\label{momentum}
Throughout this section the denotation of initial data, $(p,s)$, 
and final data, $(x,t)$, is
\begin{eqnarray}
\mathbf{d}_p&\equiv\left[(p,0),(x,t)\right],\nonumber\\
\mathbf{b}_p&\equiv\left[(p,0),(0,b_p)\right],\nonumber\\
\mathbf{r}_p&\equiv\left[(-p-b_p,b_p),(x,t)\right].\nonumber
\end{eqnarray}

The general classical path and the corresponding momentum connecting
the initial data $\left(p,s\right)$ with final data
$\left(x,t\right)$ are\numparts
\begin{eqnarray}
q(\tau;\mathbf{x}_p)=
\left(\tau-t\right)^2+2\left(\tau-t\right)
\left(p+t-s\right)+x\label{xpqgen}\\
p(\tau;\mathbf{x}_p)= \left(\tau-s\right)+p.\label{xppgen}
\end{eqnarray}
\endnumparts
The general Hamiltonian of the classical system is therefore
\begin{eqnarray}
H(\tau;\mathbf{x}_p)= \left(t-s+p\right)^2-x.\label{xpHgen}
\end{eqnarray}
Setting $s=0$ corresponds to the direct paths under consideration 
in this paper, and to the initial segments of bounce 
paths:\numparts
\begin{eqnarray}
q(\tau;\mathbf{d}_p)=\left(\tau-t\right)^2+2(\tau-t)\left(p+t\right)+x,
\label{xpq}\\
p\left(\tau;\mathbf{d}_p\right)=\tau+p\label{xpp}.
\end{eqnarray}
\endnumparts
The Hamiltonian for the direct case is therefore given by
\begin{eqnarray}
H_{p,d}=\left(p+t\right)^2-x=p^2+2pt+t^2-x,\label{Hpd}
\end{eqnarray}
and the energy of the trajectory may therefore be categorized
by:\numparts
\begin{eqnarray}
H_{p,d}>0:\quad 
p\in\left(-\infty,-\sqrt{x}-t\right)\cup\left(\sqrt{x}-t,\infty\right),
\label{Hpd>0}\\
H_{p,d}=0:\quad p=\pm\sqrt{x}-t,\label{Hpd=0}\\
H_{p,d}<0:\quad 
p\in\left(-\sqrt{x}-t,\sqrt{x}-t\right).\label{Hpd<0}
\end{eqnarray}
\endnumparts
Using equation (\ref{xpqgen}), the bounce paths may be characterized
as:\numparts
\begin{eqnarray}
q\left(\tau;\mathbf{b}_p\right)&=
\left(\tau-b_p\right)^2+2\left(\tau-b_p\right)\left(p+b_p\right),
\label{xpbq1}\\
q\left(\tau;\mathbf{r}_p\right)&=\left(\tau-t\right)^2
+2\left(\tau-t\right)
\left(t-p-2b_p\right)+x.\label{xpbq2}
\end{eqnarray}
\endnumparts
The Hamiltonian for the bounce trajectories is thus
\begin{eqnarray}
H_{p,b}=p(b_p;\mathbf{b}_p)^2=\left(p+b_p\right)^2.
\end{eqnarray}
Note that the bounce Hamiltonian is always greater than or equal to
zero, with equality for $p=-b_p\,$.

Since all trajectories are required to stay on the positive side of
the ceiling, the initial momentum for the non-bounce trajectories is
constrained by
\begin{eqnarray}
q(0;\mathbf{d}_p)=-t^2-2pt+x\geq0\nonumber.
\end{eqnarray}
Therefore a constraint on type (i) and types (ii) and (iii) 
trajectories is,
respectively,
\begin{eqnarray}
0< p\leq\frac{x}{2t}-\frac{t}{2}\,,\label{ca}\\
p\le\frac{x}{2t}-\frac{t}{2}<0\label{cb}.
\end{eqnarray}

\subsection{Trajectories of type (i)}
For the particle to initially move away from the ceiling, $p>0$.
Equation (\ref{ca}) is the only other constraint on the system
and therefore
specifies the interval of integration over $p$ for this class of 
paths.
Note that (\ref{ca})
implies $\sqrt{x}>t$; therefore, since 
$\sqrt{x}-t<\frac{x-t^2}{2t}\,$,
equations (\ref{Hpd>0}) and (\ref{Hpd<0}) imply that the energy of
type (i) trajectories may be either greater than or less than 
zero:
\begin{eqnarray}
H<0:\quad p\in\left(0,\sqrt{x}-t\right)\nonumber\\
H>0:\quad p\in\left(\sqrt{x}-t,\frac{x-t^2}{2t}\right).\nonumber
\end{eqnarray}

\subsection{Trajectories of type (ii)} Besides the requirement 
that
$p<0$, and equation (\ref{cb}), the particle must also not turn
around nor enter the forbidden region. Equation (\ref{xpp}) 
shows
that a turning point of the trajectory will occur at
$\tau=-p=\abs{p}$, therefore $t<\abs{p}$. Forbidding the 
particle to
 turn around and requiring $x>0$ are enough to ensure that the
trajectory will not enter the forbidden region. Since
$-t<\frac{x}{2t}-\frac{t}{2}$, the second constraint is weaker 
than the first and hence the operative constraint on the initial
momentum is
\begin{eqnarray}
p<-t\label{pii}
\end{eqnarray}
for trajectories of type (ii). 

Finally, equations (\ref{Hpd>0}) and (\ref{Hpd<0}) reveal that the
energy of the trajectory may again be greater than or less than
zero according to:
\begin{eqnarray}
H_{p,d}>0:\quad p\in\left(-\infty,-\sqrt{x}-t\right),\nonumber\\
H_{p,d}<0:\quad p\in\left(-\sqrt{x}-t,-t\right).
\end{eqnarray}

\subsection{Trajectories of type (iii)} Using equations \eref{n}  
and (\ref{Hpd}),
the position and energy of the particle at the turning point are
\begin{eqnarray}
q(n;\mathbf{d}_p)=x-(p+t)^2=-H_{p,d}.\label{quad}
\end{eqnarray}
Therefore, a turning-point trajectory implies that the energy is
less than zero; otherwise the turning point is behind the 
ceiling.
Using equation (\ref{Hpd<0}), the initial momentum is therefore
constrained by:
\begin{eqnarray}
-\sqrt{x}-t<p<\sqrt{x}-t\nonumber.
\end{eqnarray}
Also, the opposite of equation (\ref{pii}) must be true; 
otherwise
the turning point would not have time to occur:
\begin{eqnarray}
t>\abs {p}\Rightarrow p>-t\label{p3}.
\end{eqnarray}
If $\sqrt{x}>t\Rightarrow x>t^2$, then the upper bound in equation
(\ref{Hpd<0}) will be positive. Therefore, the regions corresponding
to the momentum of the particle and the energy being less than zero
are:
\begin{eqnarray}
x<t^2:\quad p\in\left(-\sqrt{x}-t,\sqrt{x}-t\right)\nonumber\\
x>t^2:\quad p\in\left(-\sqrt{x}-t,0\right)\nonumber.
\end{eqnarray}
However, equation (\ref{p3}) implies that the lower bounds must be
replaced by $-t$, therefore the correct constraints on the momentum
are:\numparts
\begin{eqnarray}
x<t^2:\quad p\in\left(-t,\sqrt{x}-t\right),\\
x>t^2:\quad p\in\left(-t,0\right).
\end{eqnarray}
\endnumparts

\subsection{Bounce trajectories} Since the construction of
equations \eref{xppgen}, \eref{xpbq1}, and \eref{xpbq2} 
guarantees the validity of
(\ref{pc}), the final constraint to impose on the bounce
trajectories is the location of the ceiling:
\begin{eqnarray}
q(b_p;\mathbf{r}_p)=0.\label{xpbbc}
\end{eqnarray}
This yields the quadratic equation
\[
b_p^2+b_p\left(\frac{2}{3}p-\frac{4}{3}t\right)+\frac{1}{3}
\left(t^2-2pt-x\right)=0,\]
with solution
\[b_p=\frac{1}{3}\left(2t-p\pm\sqrt{\left(p+t\right)^2+3x}\right).
\]
Requiring the time of bounce, $b_p\,$, to be less than the 
trajectory time, $t$, and
the initial momentum to be less than zero yields
\begin{equation}\label{bounce constraint}
\left(\abs{p}-t\right)\pm\sqrt{\left(\abs{p}-t\right)^2+3x}\,\leq0
,\end{equation}
which requires the negative root.
Therefore, independently of the relationship between the initial
momentum and the trajectory time, the correct bounce time is given
by
\begin{eqnarray}
b_p=\frac{1}{3}\left(2t-p-\sqrt{\left(p+t\right)^2+3x}\right).
\label{bouncetimep}
\end{eqnarray}
Constraining the initial position of the trajectory to be in the
classical region implies
\begin{equation}
q(0;\mathbf{b}_p)> 0\Rightarrow
0<b_p<-2p.\label{b3}\end{equation}
However, the ``critical'' trajectory for the bounce path is when the
initial momentum is equal to the time of bounce, which corresponds
to the particle just grazing the ceiling. All other initial 
momenta
must be greater, in magnitude, than the time of bounce:
\begin{eqnarray}
0<b_p<-p.
\label{b4}\end{eqnarray}
This constraint is stronger than \eref{b3} and 
therefore is the final constraint to impose on the bounce 
trajectories.
Equality on the right side of \eref{b4} implies 
\begin{equation}
2t+2p-\sqrt{\left(t+p\right)^2+3x}=0.
\label{tempconstraint}\end{equation}
This leads, on the one hand, to
\begin{equation}
p=-t\pm\sqrt{x}\label{pbounceconstraint}.
\end{equation}
But now \eref{tempconstraint} also requires that $t+p>0$, so the 
correct solution is
\begin{eqnarray}
p=\sqrt{x}-t.\label{p2}
\end{eqnarray}
On the other hand, the initial momentum
corresponding to a bounce time equal to zero is the
solution of
\[
2t-p-\sqrt{\left(t+p\right)^2+3x}=0,\]
or
\begin{equation}
p=\frac{t^2-x}{2t}\,.\label{p1}
\end{equation}
The right-hand sides of \eref{p1} and \eref{p2} have opposite 
signs, and only negative values of them are operative.  
Therefore, 
the constraints to impose on the initial momentum so
that both inequalities (\ref{b4}) hold are 
\numparts
\begin{eqnarray}
x<t^2:\quad p<\sqrt{x}-t,\\
x>t^2:\quad p<\frac{t^2-x}{2t}\,.
\end{eqnarray}
\endnumparts

The  constraints on the initial momentum for given final
data $\left(x,t\right)$ are summarized in Table \ref{tablep}. 
They are somewhat more complex than those for given initial 
position in \sref{position}, although in compensation we did not 
need to solve a cubic equation this time.

\Table{\label{tablep} Constraints on the initial momentum given 
the
final data $\left(x,t\right)$. Endpoint cases are not included.}
\br $x\in$
&$p\in$&Trajectory\\
\mr
$\left(t^2,\infty\right)$&$\left(0,\frac{x-t^2}{2t}\right)$&Type 
(i) (rightward)\\
$\left(0,\infty\right)$&$\left(-\infty,-t\right)$&Type (ii) 
(leftward)\\
$\left(0,t^2\right)$&$\left(-t,\sqrt{x}-t\right)$&Type 
(iii) (turning)\\
$\left(t^2,\infty\right)$&$\left(-t,0\right)$&Type (iii) 
(turning)\\
$\left(0,t^2\right)$&$\left(-\infty,\sqrt{x}-t\right)$&Bounce\\
$\left(t^2,\infty\right)$&$\left(-\infty,\frac{t^2-x}{2t}\right)$&Bounce\\
\br
\endTable

If $x\in(t^2,\infty)$, 
all four types of trajectories may occur for various initial 
momenta.

If the initial momentum is 
positive, only the trajectory of type (i) is allowable. 
Even it does not exist if $p$ is too large for the given $(x,t)$
(or if $x<t^2$).

For $p<0$, further
analysis reveals a distinction between  the regions
  $x\in(t^2,3t^2)$ and
   $x\in(3t^2,\infty)$.
In both cases the monotonic (type (ii)) and 
turning regimes of $p$ are mutually exclusive.
When $x<3t^2$, 
 the bounce 
regime overlaps both of those, 
but when $x>3t^2$,
 the bounce regime is a subset of the monotonic one.
In either case there is a possibility of only one path --- that 
is, a bounce path may not exist for the given data.

For $x\in(0,t^2)$ there are no paths corresponding to positive
initial momentum. The intervals corresponding to the always 
left-moving  and the
turning trajectories are disjoint and 
their union coincides with the bounce interval.
Therefore, two paths are always 
possible for $x\in(0,t^2)$;
one will always be a bounce, and the other is either type (ii)
$(p<0)$ or  turning.

In short,  
 there are regimes of $(p,x,t)$ data where only one path  
exists and other regimes where two exist,  or none.
In contrast, for $(y,x,t)$ data, if a  path exists at all 
  (i.e., \eref{tbound} is satisfied), 
 then there will always be two distinct trajectories,
except in the very special case of a critical trajectory.

\subsection{The action}\label{xpaction}
Using equations (\ref{xpqgen}) and (\ref{xppgen}), the classical
Lagrangian for the general initial and final conditions is found to be
\begin{eqnarray}
L_p(\tau;\mathbf{x}_p)=
\left(\tau-s+p\right)^2
+\left(\tau-t\right)^2+2\left(\tau-t\right)\left(p+t-s\right)+x.
\label{plagr}\end{eqnarray}
Thus the general action for $\tau=t$ and arbitrary
initial and final data, $\mathbf{x}_p\,$, up to a constant of
integration, $S_0$, is
\begin{eqnarray}
S_p(\mathbf{x}_p)=
-\,\frac{\left(t-s\right)^3}{3}+\left(t-s\right)\left(x+p^2\right)
+S_0 = S+S_0\,.
\label{xpSgen}
\end{eqnarray}
The following partial derivatives of the general action will again
prove useful in determining the appropriate initial
condition:\numparts
\begin{eqnarray}
\frac{\partial S}{\partial t}(\mathbf{x}_p)
=-\left(t-s\right)^2+\left(p^2+x\right),\label{dxpSgendt}\\
\frac{\partial S}{\partial
s}(\mathbf{x}_P)=\left(t-s\right)^2-\left(p^2+x\right)
= -\,\frac{\partial S}{\partial t}\,,\label{dxpSgendt0}\\
\frac{\partial S_p}{\partial
x}(\mathbf{x}_p)=\left(t-s\right).\label{dxpSgendx}
\end{eqnarray}
\endnumparts
Using the derivatives of the general initial position from equation
 (\ref{xpqgen}),
\numparts
\begin{eqnarray}
\frac{\partial q}{\partial
t}(s;\mathbf{x}_p)=-2\left(p+t-s\right),\label{dxpqgendt}\\
\frac{\partial q}{\partial x}(s;\mathbf{x}_p)=1\label{dxpqgendx},
\end{eqnarray}
\endnumparts
one sees that the correct initial action is
\begin{eqnarray}
S_0=p(s;\mathbf{x}_p)\,q(s;\mathbf{x}_p)=p\,q(s;\mathbf{x}_p),\nonumber
\end{eqnarray}
because then (cf.~\eref{Sreversed})
\numparts\begin{eqnarray}
 \frac{\partial S_p}{\partial t} = -H(\mathbf{x}_p), \\
 \frac{\partial S_p}{\partial x} = p(t;\mathbf{x}_p), \\
 \frac{\partial S_p}{\partial p} = q(s;\mathbf{x}_p).
\end{eqnarray}\endnumparts
Then the actions for the direct and bounce cases
are respectively\numparts
\begin{eqnarray}
\fl S_{pd}&=&S_p(\mathbf{d}_p)+ p\,q(0;\mathbf{d}_p)\nonumber\\
\fl &=&-\frac{t^3}{3}-p\,t(p+t)+x(p+t),\label{xpSd}\\
\fl S_{pb}&=&S_p(\mathbf{b}_p)+S_p(\mathbf{r}_p)+
 p\,q(0;\mathbf{b}_p)\nonumber\\
\fl&=&-\frac{1}{3}\left[b_p^3+\left(t-b_p\right)^3\right]
-b_pp\left(p+b_p\right)+
\left(t-b_p\right)\left(x+\left(p+b_p\right)^2\right)
-pb_p(2p+b_p).\label{xpSb}
\end{eqnarray}
\endnumparts
In these calculations, terms resulting from the differentiation of $b_p$
play a crucial role,
whereas in the initial-position formulation they cancel each other.

\subsection{The amplitude}

The Jacobian corresponding to the amplitude function is
$\abs{\frac{\partial q\left(s\right)}{\partial q\left(t\right)}}$.
Using equation (\ref{xpqgen}) for the trajectories yields for the
direct case
\begin{eqnarray}
\frac{\partial q(0)}{\partial x}=1,\label{xpdjacobian}
\end{eqnarray}
which reveals that the amplitude for the direct case is a constant
solely depending on the initial density of particles. This initial
density must be in agreement with the the initial form of the
quantum propagator in the momentum representation, and therefore 
\begin{eqnarray}
A_d=\frac{1}{\sqrt{2\pi}}\,.\label{xpampdir}
\end{eqnarray}
Using equations (\ref{xpbq1}) and (\ref{xpbq2}) for the
trajectories in the Jacobian for the bounce amplitude yields
\begin{eqnarray}
\frac{\partial q_1(0)}{\partial
x}=\frac{p+b_p}{\sqrt{\left(p+t\right)^2+3x}}\,,
\end{eqnarray}
and therefore the amplitude function for the bounce trajectory is,
up to a phase,
\begin{eqnarray}
A_b=\frac{1}{\sqrt{2\pi}}\sqrt{\abs{\frac{p+b_p}
{\sqrt{\left(p+t\right)^2+3x}}}}\,,\label{Apb}
\end{eqnarray}
where the initial (normalization) factor is carried over from the
incident direct path.

Since  the critical curve for the bounce trajectory is
equivalent to the $E=0$ type-(iii) trajectory, it is evident that
$b_p=n_p=-p$ for the critical trajectory. Hence the amplitude will
vanish for the critical path. All other values of $b_p$ are less
than $-{p}$, by \eref{b4}.
 Therefore, the numerator of equation (\ref{Apb}) will
be negative for bounce paths.
As in \sref{position}, our phase convention is
 that the direct and bounce propagators
 agree at the ceiling, $x=0$.
Therefore, the correct  bounce amplitude is
\begin{equation}
A_{pb}=\frac{1}{\sqrt{2
\pi}}\sqrt{\frac{|p+b_p|}{\sqrt{(p+t)^2+3x}}}\,.
\label{Apb2}\end{equation}
And, with this convention, 
to $O(\hbar)$ the WKB approximation for the quantum
propagator is the difference,
$U_{p\mathrm{d}} - U_{p\mathrm{b}}\,$.
of the propagators constructed from
the bounce and non-bounce paths:\numparts
\begin{eqnarray}
\fl U_{p\mathrm{d}}(x,p,t)&=\frac{1}{\sqrt{2\pi}}\,
\exp\left[\rmi\left(-\frac{t^3}{3}+t\left(p^2+x\right)
-p\left(t^2+pt-x\right)\right)\right],\label{Upd} \\
\fl
U_{p\mathrm{b}}(x,p,t)&=-\,\frac{1}{\sqrt{2\pi}}\sqrt{\frac{|p+b_p|}
{\sqrt{\left(p+t\right)^2+3x}}}\,
\exp\left[i\left(-\frac{1}{3}\left[b_p^3+\left(t-b_p\right)^3\right]
-b_pp\left(3p+2b_p\right)\right)\right]\nonumber\\
&\times\exp\left[i\left(\left(t-b_p\right)
\left(x+\left(p+b_p\right)^2\right)\right)\right].\label{Upb}
\end{eqnarray}
\endnumparts 
Here $b_p$ is defined by \eref{bouncetimep}.

\section{Numerical comparison of the propagators}  
\label{numerical}

In this section we try out the propagators 
 (\ref{Upd}), (\ref{Upb}), (\ref{Uyd}), and (\ref{Uyb}) 
by applying them to localized wave packets.
We take 
 the initial state of the
system in position space to be a general Gaussian wave packet:
\begin{equation}
\psi\left(y\right)=\left(\frac{2}{\gamma\pi}\right)^{\frac{1}{4}}
\rme^{-\frac{(y-\bar{y})^2}{\gamma}-\rmi y\bar{p}},\label{initialq}
\end{equation}
where $y$ denotes the initial position, the average initial position
is $\bar{y}$, the average initial momentum is $\bar{p}$, and the
constant $\gamma$ prescribes the width of the packet. 
Then the initial wave packet in momentum space is the 
Fourier transform of \eref{initialq}:
\begin{equation}
\phi\left(p\right)=\left(\frac{\gamma}{2\pi}\right)^{\frac{1}{4}}
\rme^{-\gamma\frac{(p+\bar{p})^2}{4}-\rmi
\bar{y}(p+\bar{p})},\label{initialp}
\end{equation}
which is also Gaussian, with width $4/\gamma$.

Gaussians are chosen to make the treatment as symmetrical as 
possible between position space and momentum space, thereby 
minimizing any bias in numerical calculations and their 
interpretation.    Unfortunately, the support of the Gaussian 
$\psi(y)$ extends into the forbidden region $y<0$, 
so it does not represent a state of our system, strictly 
speaking.
A momentum-space counterpart of this fact  is that  
the signed momentum does not exist as a legitimate 
quantum observable (a self-adjoint operator) when the 
configuration space is a half-line.  An exact Fourier analysis of 
such a system would require Fourier sine transforms and a 
``momentum'' observable that is nonnegative.  However, the whole 
spirit of the semiclassical approximation requires that the 
particle be thought of as approximately localized and having an
approximate classical momentum, and our foregoing calculations 
have been conducted in this framework.
If
 \begin{equation}
\bar{y}\gtrsim \gamma, 
\label{goodgamma}
\end{equation}
then we expect the semiclassical picture to hold.
Furthermore, because of the rapid decay of the Gaussian function, 
whenever the wave packet $\psi(y)$ or $\phi(p)$ is located well 
inside a classically allowed region in the sense of Table
\ref{tablex} or~\ref{tablep}, the limits of integration 
can be extended to infinity without committing great error.

\begin{figure}
\centering \caption{Plot of WKB evolution of an initial wave 
packet, from the point of view of a final point $(x,t)$
The solid lines are the classical trajectories to that point for 
a particle initially 
located at the average value of the initial wave packet;  the
dashed lines are the classical paths starting at points
rather far into the wings of the initial packet.
 The classically forbidden region for trajectories is at the 
lower left of the plot, bounded by the critical trajectory (not 
shown), which intersects the horizontal axis at a point $y_c$
(compare Figure \ref{bounce}, inverted).}
\includegraphics{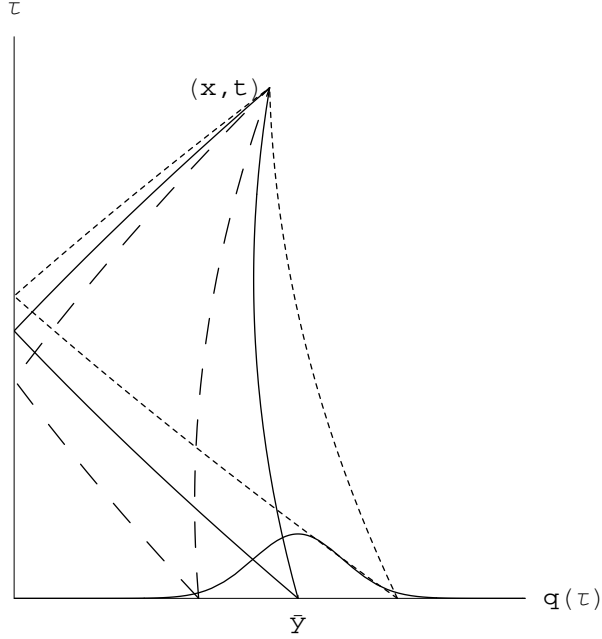}
\label{packet}
\end{figure}
Figure \ref{packet} depicts the evolution of an initial Gaussian
wave function from an initial time $\tau=0$ to a final point
$(x,t)$. From each initial point to the right of a critical 
value, $y_c\,$, there are two trajectories that arrive at $(x,t)$.
The contribution of each path to the wave function at $(x,t)$
is weighted by the initial wave packet and otherwise determined 
by the propagator (direct or bounce, as appropriate).
The value of $y_c$ is determined from \eref{tbound} to be
\begin{equation}
y_c = (t-\sqrt{x})^2.
\label{yc}\end{equation}
Note that the roles of $x$ and $y$ in Figures \ref{bounce} and 
\ref{packet} are reversed, so that in Figure \ref{packet} the 
classically forbidden 
region is below and to the left of the critical trajectory 
starting from $y_c\,$.

\begin{figure}
\centering \caption{Comparison of the evolution of initial wave
functions  by the WKB propagators associated with the
classical direct paths. The initial data is such that $\gamma=2$ 
and
$\bar{p}=-6$. The final data is $(x,t)=(4,5)$, so that $y_c=9$. 
The average initial position, $\bar{y}$, is varied in the plot.}
\includegraphics{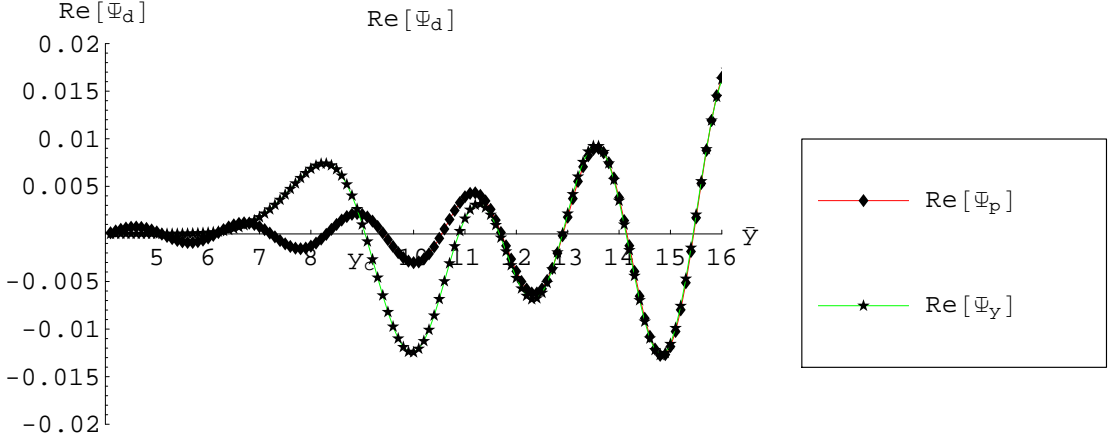}
\label{dircomp}
\end{figure}

\begin{figure}
\centering \caption{Comparison of the evolution of initial wave
functions  by the WKB propagators associated with the
classical bounce paths. Again, 
$\gamma=2$, 
$\bar{p}=-6$,  $(x,t)=(4,5)$, $y_c=9$, and the average 
initial position, $\bar{y}$, is the independent variable.}
\includegraphics{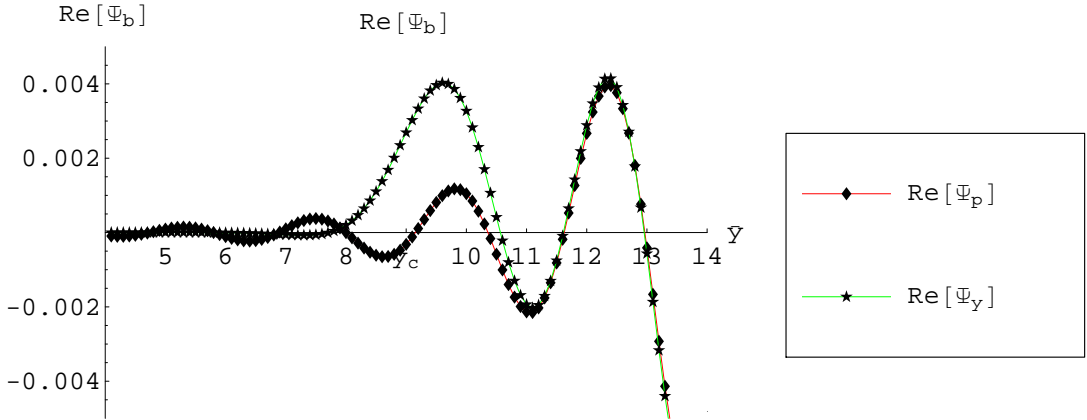}
\label{boucomp}
\end{figure}

 Figures \ref{dircomp} and \ref{boucomp} 
show the real parts of the direct and bounce contributions 
to the final wave function at the  point $(x,t)=(4,5)$.
From Table \ref{tablep} one can see that the classical limits on 
the initial momentum 
 data are $(-\infty,-3)$.
(The significance of $p=-3$ is that it is the critical value 
where $b_p=-p\,$; see \eref{bouncetimep}.)
The classical limits on initial position are 
 $(9,\infty)$, in accord with \eref{yc}.
We choose $\gamma=2$, so that  the width of the packet in 
natural units is the same in both $y$ and $p$ space,
and  the effective supports of the initial 
wave packets (encompassing $\sim 95\%$ of
the  packet) for the position and momentum cases are 
$(\bar{y}-2,\bar{y}+2)$ and
$(\bar{p}-2,\bar{p}+2)$, respectively. 
Thus (a) \eref{goodgamma} is well satisfied for $\bar{y} \ge 4$;
(b) the initial momentum-space packet is well inside the 
classically allowed region if $\bar{p}=-6$ (as we arbitrarily 
choose for the plots); (c) the initial position-space packet is 
well inside the classically allowed region if $\bar{y} \ge 11$.
The computations indeed show that the two propagators give 
essentially equal results for $\bar{y} \ge 11$ but not for 
smaller $\bar{y}$. 
 Our interpretation then is that the 
momentum-space calculation should be preferred for 
$4<\bar{y}<11$, and neither should be trusted for smaller 
$\bar{y}$.
(The position-space solution might become superior when $\bar{p}$
is too close to an endpoint of a classically allowed interval of 
momentum, but we have not verified that.)
In Figures \ref{dircomp} and \ref{boucomp} the two 
pieces of the position-space 
solution make  spurious large excursions in the critical region
$7<\bar{y}<11$ and then fall rapidly to $0$ for smaller 
$\bar{y}$, as expected, since the $y$-space propagator is 
identically $0$ in that region and the wave function is coming 
entirely from a wing of the Gaussian packet.
The more trustworthy momentum-space solution is small but 
nontrivial in that region, again as expected.

\ack

We thank B.-G. Englert for help in solving the cubic equation 
\eref{cubic1} and S.-A. Chin for asking, ``Why doesn't the method 
of images work?'' This research has been supported in part by 
National Science Foundation Grants PHY-0554849 and PHY-0968269.

\appendix
\section{Why the method of images does not solve the problem}

 It is natural to expect that a Green function for a problem with a 
flat, perfectly reflecting boundary can be constructed from the 
Green function for all of space by subtracting the Green function 
for an image source located symmetrically on the other side of the 
boundary. When a potential function is involved, however, the 
situation is not as simple as it seems, especially when the 
derivative of the potential is not zero at the boundary.

  In our problem \eref{schrod}, it is easy to fall into either of 
two traps.

 First, we know the ``free'' propagator \eref{freeprop}.   It might 
seem that
 $U_\mathrm{free}(t,x,y) - U_\mathrm{free}(t,x,-y)$
 is the propagator $U(t,x,y)$ for the scenario with the ceiling 
present, as it would be if the potential were absent.
 But this function does not satisfy the boundary condition, 
$U(t,0,y)=0$.  A source ``uphill'' does not have the same effect at 
the boundary as a source ``downhill''\negthinspace.

 The second variant of the fallacy is to write
$U_\mathrm{free}(t,x,y) - U_\mathrm{free}(t,-x,y)$.
 This time the boundary condition is satisfied, but the second term 
obeys the wrong differential equation (as a function of $t$ 
and~$x$), because the sign of the potential has been reversed.

 The only correct way to apply the method of images is \cite{AS}
 to extend the potential to negative~$x$ as an even function:
 $V(x) = -|x|$.  With the modified potential, 
$U_\mathrm{free}(t,x,y)$ will be invariant under simultaneous sign 
change of $x$ and~$y$, so that the two formulas proposed previously 
are equivalent and either  satisfies the propagator problem.
(The reflected term in the propagator presumably 
has as its semiclassical approximation 
the WKB contribution of the bounce paths.)
Although indisputably correct, this construction is useless for our 
purposes, because solving the Schr\"odinger equation for the 
piecewise defined potential, $|x|$, is at least as hard as solving 
the original ceiling problem.
 Indeed, the standard textbook advice for solving a problem on the 
whole real line with an even potential is to decompose into odd and 
even modes, and to find the latter by solving the problem on the 
half-line with Dirichlet and Neumann boundary condition, 
respectively.  Instead of solving the ceiling problem, the gambit 
has doubled~it.

\end{document}